# Bending-induced extension in two-dimensional crystals

Douxing Pan,[1,3][†] Yao Li,[2][†] Tzu-Chiang Wang,[1],* and Wanlin Guo[2],*

[1]State Key Laboratory of nonlinear Mechanics, Institute of Mechanics, Chinese Academy of Sciences, Beijing 100190, China
[2]State Key Laboratory of Mechanics and Control for Mechanical Structures and Key Laboratory for Intelligent Nano Materials and Devices (MOE), Nanjing University of Aeronautics and Astronautics, Nanjing 210016, China
[3]University of Chinese Academy of Sciences, Beijing 100049, China
(Dated: October 6, 2015)

According to the classical theory of elasticity, a plate subjected to a bending moment always deflects with symmetric tensile and compressive strains in its two sides, without overall deformation perpendicular to the bending moment. Here, we find by *ab initio* simulations that significant overall tensile strain can be induced by pure bending in a wide range of two-dimensional crystals perpendicular to the bending moment, just like an accordion being bent to open. This accordion effect is raised by asymmetric response of chemical bonds and electron density to the bending curvature, with the tensile strain being a power function of the curvature.

PACS number: 62.20.-x, 61.48.-c, 46.90.+s, 81.07.-b

Graphene and graphene-like two-dimensional (2D) crystals have rapidly raised in the past decade as the most promising materials with exceptional physical behaviors [1-2]. The most unique characters of the 2D crystals are their atomic thinness and extremely high flexibility [2,3]. Although 2D crystals exhibit very excellent in-plane mechanical properties and the in-plane uniaxial and isotropic biaxial strain has difficulty to open the band gap in graphene [4], it has been widely known that their electronic and optical performances are extremely sensitive to flexion deformation [5-10]. For example, graphene is regarded as the strongest material in the world with a Young's modulus of 1 terapasal [11] and the Young's modulus of monolayered $MoS_2$ is comparable to that of steel [12], but they are extremely easy to bend, ripple and even can be directly rolled into nanotubes [13]. Most recently, an interesting bending Poisson effect is revealed in a series of 2D crystals where bending deformation is found to be able to induce deformation along the bending moment [14]. The 2D monolayers are not only flexible, they are also very stable under serious bending deformation, or they can be seriously folded without breaking. The graphene and h-BN monolayers can be stable to curvature radius down to about 0.15 nm [15,16] and the black phosphorus (BP) monolayer can be stable to a curvature radius of 0.5 nm when bending along zigzag and much smaller along armchair direction [17]. However, the detailed bending behaviors, especially the coupling between flexion and in-plane deformation in 2D crystals remains in poor understanding and has never been reported in many 2D crystals, although some interesting coupling between in-plane stretching and bending or twisting behaviors has been reported in graphene and its ribbon structures [18-20]. In this letter, we show by *ab initio* simulations that bending 2D crystals can induce significant overall tensile deformation perpendicular to the bending moment, just like an accordion being bent to open. The underlying mechanism is revealed through comprehensive investigations of the responses of chemical bonds and electron density to bending curvature. Interestingly, such accordion effect is found to be intrinsic to all the 2D crystals investigated in this work.

In the following, the results of BP monolayers are used to show how the accordion effect occurs before the intrinsic accordion effect in a wide range of 2D crystals being presented. The geometry relaxations are carried out within the framework of density-functional theory (DFT) [21], as implemented in the VASP code [22,23]. The generalized gradient approximation in the form of the Perdew-Burke-Ernzerhof exchange correlation functional [24] and projector augmented wave potentials are employed [25-27]. High accuracy settings are adopted in the simulations [28].

The relaxed lattice constants of the BP unit cell are $a_1$ = 0.4620 nm, $a_2$ = 0.3299 nm, and the length $L_0$ of a planar strain-free BP monolayer can be written as

$$L_0 = \left( n^2 |a_1|^2 + nm |a_1||a_2| + m^2 |a_2|^2 \right)^{\frac{1}{2}}, \qquad (1)$$

where *n* and *m* are the number of BP unit cells along the armchair and zigzag directions, respectively. To mimic the bending deformation, BP monolayers are rolled into the corresponding closed black phosphorus nanotubes (PNTs) [14] and the reciprocal of the radius is defined as the bending curvature, as shown by Fig. 1. When the BP monolayer is rolled into an armchair PNT (aPNT), it is called armchair bending, or arm-bending; when rolled into a zigzag PNT (zPNT), it is called zigzag bending, or zig-bending.

Based on DFT simulation results, the tensile and compressive strains induced by pure bending in the outer and inner half-layers of BP are defined as,

$$\varepsilon_t = \frac{L_{out} - L_0}{L_0} \quad \text{and} \quad \varepsilon_c = \frac{L_{in} - L_0}{L_0}, \qquad (2)$$

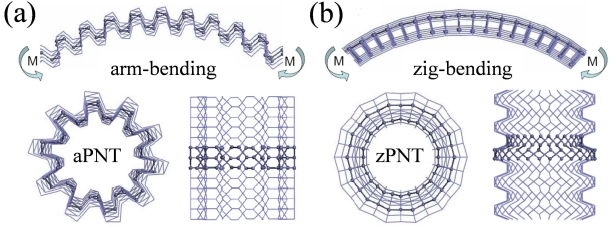

FIG. 1 (color online). The monolayered BP in bent state. (a) Arm-bending with the bending moment M being perpendicular to the armchair direction, and an armchair BP nanotube (aPNT). (b) Zig-bending with the bending moment M being perpendicular to the zigzag direction, and a zigzag BP nanotube (zPNT).

respectively, where $L_{out}$ and $L_{in}$ are the length of the outer and inner half-layers of the BP monolayer in the bent state as defined in Fig. S1 in the Supplementary Material [29]. The tensile strain of the mid-plane of the bent BP monolayer can be defined as

$$\varepsilon_m = \frac{\varepsilon_t + \varepsilon_c}{2}, \qquad (3)$$

which presents the overall tensile deformation due to bending. In this work, the curvature means the curvature of the mid-plane of the bent monolayer.

Figure 2 shows that bent BP monolayer exhibits apparent tensile strain in its outer half-layer and compressive strain in its inner half-layer perpendicular to the bending moment or the axial direction of the PNT. However, the tensile strain within the outer half-layer is significantly larger than the compressive strain within the inner half-layer [Figs. 2(a) and 2(b)], resulting in a significant overall tensile deformation perpendicular to the bending moment [Fig. 2(c)]. This bending induced overall extension is impossible in the classical theory of elasticity [30], where a plate subjected to a pure bending moment always deflects symmetrically with tensile strain in its outer side and equal compressive strain in its inner side. Such an asymmetric deformation is similar to an accordion being bent to open.

Comparison of Fig. 2(a) and Fig. 2(b) shows that the zig-bending induces a larger extension and smaller compression than arm-bending. For zig-bending, the change of tensile and compressive strain with bending curvature $k$ exhibits more nonlinearity. The compressive strain even turns into decrease with further increasing curvature beyond 1 nm$^{-1}$. It shows that the accordion effect in BP monolayer is anisotropic with the mid-plane strain under zig-bending being 2.27 times of that under arm-bending at fixed bending curvature of 1.36 nm$^{-1}$.

When the bending curvature is down to about 0.5 nm$^{-1}$, the tensile and compressive strains are $\varepsilon_t$ = 6.26%

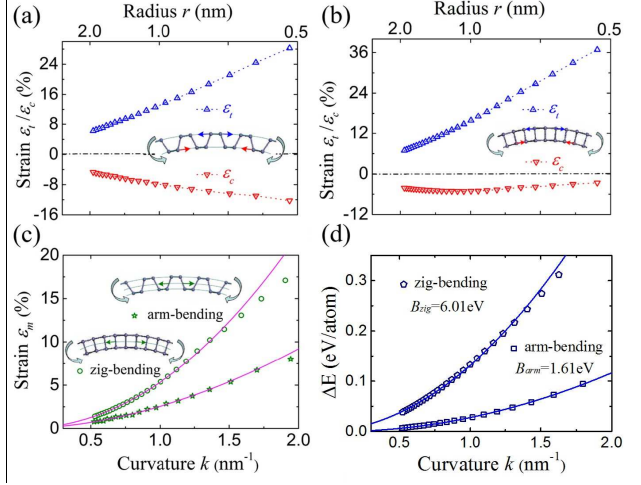

FIG. 2 (color online). Bending induced extension. (a) Arm-bending and (b) zig-bending induced tensile and compressive strains in the outer and inner half-layers of BP as a function of curvature. (c) The tensile strain of mid-plane in BP in the bent state. (d) Bending strain energy, from which the bending moduli can be obtained. The solid curves in (c) and (d) are power-law and quadratic fitting of the DFT results for $k <$ 1.5 nm$^{-1}$, respectively.

and $\varepsilon_c$ = -4.71% for arm-bending and $\varepsilon_t$ = 6.93% and $\varepsilon_c$ = -4.14% for zig-bending, respectively. At the bending curvature near 2.0 nm$^{-1}$, the tensile and compressive strains for arm-bending are $\varepsilon_t \approx$ 28.25% and $\varepsilon_c$ = -12.22%, and for zig-bending $\varepsilon_t \approx$ 36.82% and $\varepsilon_c$ = -2.62%, respectively.

Figure 3(a-c) shows the variation of bond lengths and angles in the BP monolayer with curvature under arm-bending. It can be seen from Fig. 3(b) that with increasing curvature, the bond angle $\alpha_{out}$ increases more remarkably than the decrease in $\alpha_{in}$. So, it is more difficult to compress the inner half-layer than to stretch the outer half-layer. Fig. 3(c) shows that the bond lengths $d_{out}$ and $d_m$ increase while $d_{in}$ decreases with increasing curvature. As a result, the faster increasing $\alpha_{out}$ and increase in both $d_m$ and $d_{out}$ will dramatically extend the length of the outer half-layer $L_{out}$ and the length of the mid-plane, although the decrement in $\alpha_{in}$ and $d_{in}$ will slightly shorten the length of the inner half-layer $L_{in}$.

Similarly, for the BP monolayer under zig-bending, as shown in Figs. 3(d), 3(e) and 3(f), the bond angle $\beta_{out}$ increases remarkably with increasing curvature; while $\beta_{in}$ decreases slightly, even comes into decrease with further increasing curvature beyond about 1 nm$^{-1}$. Similar trends can be seen for bond lengths $d_{out}$ and $d_{in}$ in Fig. 3(f). The non-monotonic changes of bond angle

$\beta_{in}$ and bond length $d_{in}$ lead to more significant accordion effect in zig-bending BP. As will be shown in the following, this asymmetric variation of bonding structure in bent BP monolayer can be further attributed to bending induced shift in electron density.

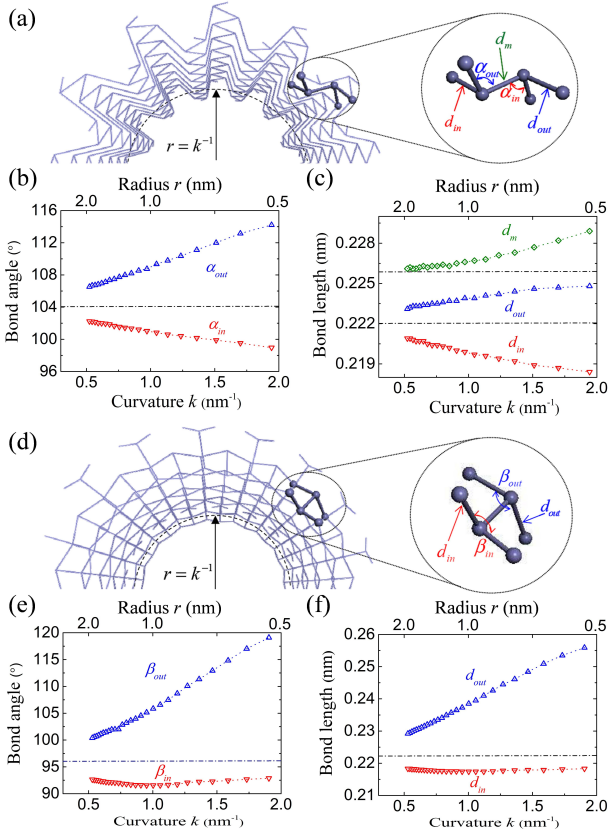

FIG. 3 (color online). Bonding structures in BP under pure bending. (a-c) Arm-bending. (a) Bonding structures. (b) Bond lengths and (c) bond angles change with bending curvature. The corresponding results for zig-bending are presented in (d-f). The dot-dash lines in (b-f) denote the corresponding values in the planar strain-free BP monolayer.

In a planar BP monolayer, the projected density of electrons (PDOE) is symmetric about the mid-plane of the monolayer as shown by Figs. 4(a) and 4(b) from both armchair and zigzag directions. Under arm- and zig-bending, the PDOE becomes asymmetric about the mid-plane, as shown by Fig. 4(c) and Fig. 4(d), respectively. The higher electron density in the compressive side will constrain the compressive strain, as the high density electrons tend to repel each other. Figs. 4(e) and 4(f) show increasingly asymmetric chemical potentials with decreasing curvature radius under arm- and zig-bending. The chemical potential in the tensile side (TP) increases with increasing curvature, while the chemical potential in the compressive side (CP) decreases with decreasing curvature radius. The induced potential differences will result in an in-built electrical field. As a result, the high electron density in the compressive side and the formation of the in-built electrical field will limit the compressive strain, so extend the length of the mid-plane, just like bending to open an accordion. Comparing the results in Fig. 4(e) and Fig. 4(f) shown that the induced potential difference is slightly larger under zig-bending than under arm-bending. This larger potential difference can explain the stronger accordion effect under zig-bending as shown in Fig. 2.

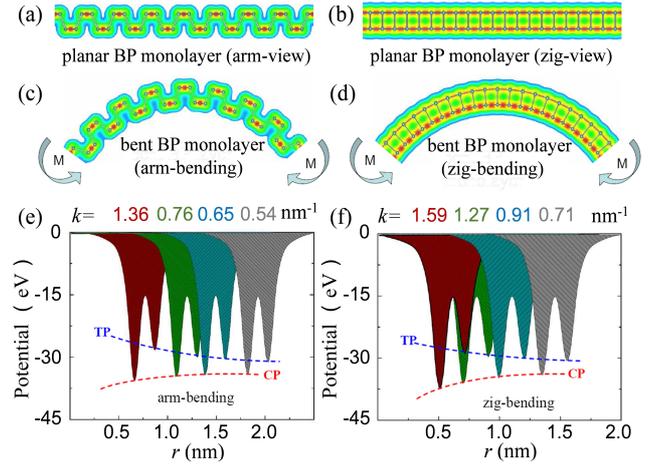

FIG. 4 (color online). Projected density of electron (PDOE) and distribution of chemical potential. (a,b) The PDOE for the planar BP monolayer viewed from (a) armchair direction and (b) zigzag direction. (c) The PDOE for BP monolayer under arm-bending and (d) zig-bending at the same bending curvature. The saturation value of the PDOE is 0.1 e/Bohr$^3$ in (a,c), and 0.065 e/Bohr$^3$ in (b,d). (e,f) Deformation potential along the thickness of the BP monolayer under bending. (e) Arm-bending and (f) zig-bending. The variation curves of the lowest potential in the tensile side (TP) and that in the compressive side (CP) with radius are drawn in (e,f).

Our density functional calculations also show that the band gap of a bent BP monolayer decreases with increasing bending curvature along both armchair and zigzag directions, with the valence band maximum raising and conduction band minimum decreasing, and the electron densities corresponding to the valence band maximum becoming asymmetric about the mid-plane in a similar way as the PDOE shown in Figs. 4(c) and 4(d) (see Fig. S3 in the Supplemental Material [29]). The energy band structures and partial charge densities corresponding to the valence band maximum of planar BP monolayer are presented in Fig. S2 in the Supplemental Material [29].

Extensive density functional theory based similar modeling and calculations show that the revealed accordion effect is intrinsic to monolayered blue phosphorus (RBP), transition metal dichalcogenides, even the single atomic thin graphene and h-BN monolayers, as shown by the results in Fig. 5 and Supplemental Material, Fig. S4 [29]. The bending induced mid-plane tensile strains in monolayered molybdenum disulfide

It is interesting that the mid-plane tensile strain $\varepsilon_m$ always increases as a simple power function of the bending curvature $k$,

$$\varepsilon_m = ck^\lambda. \qquad (4)$$

Table I. The fitted coefficient $c$ and exponent $\lambda$ in the power law $\varepsilon_m = ck^\lambda$.

| 2D crystals | | BP | MoS$_2$ | RBP | GR | BN | WS$_2$ | MoSe$_2$ | WSe$_2$ |
|---|---|---|---|---|---|---|---|---|---|
| arm-bending | $c$ | 2.55 | 2.77 | 0.39 | 0.10 | 0.08 | 2.82 | 4.24 | 4.50 |
| | $\lambda$ | 1.85 | 1.88 | 1.95 | 2.24 | 2.10 | 2.14 | 2.04 | 2.09 |
| chi-bending | $c$ | | | 0.49 | 0.10 | | | | |
| | $\lambda$ | | | 1.95 | 2.24 | | | | |
| zig-bending | $c$ | 5.33 | 3.91 | 0.64 | 0.10 | 0.14 | 4.05 | 5.79 | 5.84 |
| | $\lambda$ | 2.07 | 1.79 | 2.15 | 2.24 | 2.21 | 2.12 | 2.10 | 2.06 |

and other transition metal dichalcogenides are almost as strong as that in BP, but becomes much weaker in blue phosphorus, only about one third of that in BP. In monolayered graphene and h-BN, the bending induced tensile mid-plane strains become one order lower, where the flexion induced asymmetric π-electron density should be the only reason for the accordion effect. In all cases except graphene, the accordion effect under zig-bending is more significant than under arm-bending. Therefore, the accordion effect is closely dependent on the elements, atomic structure, chirality and thickness of the 2D crystals.

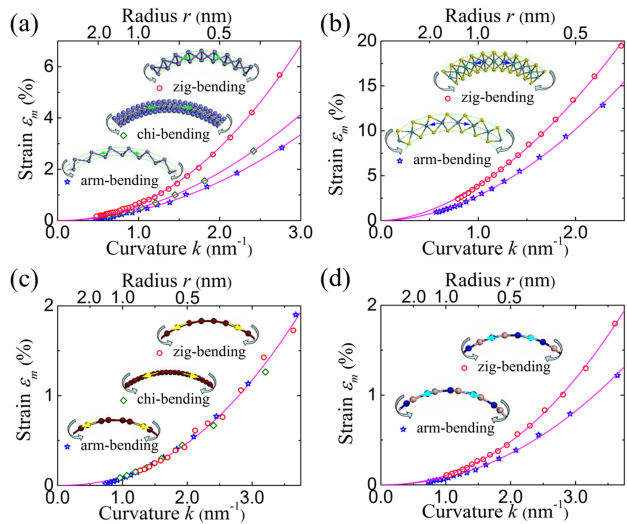

FIG. 5 (color online). Bending to tensile deformation in other two-dimensional crystals. (a) Blue phosphorus, (b) molybdenum disulfide, (c) graphene, (d) boron nitride. Here, chi-bending means the monolayers rolled up into (2n, n) nanotubes. The points are the DFT results and the solid lines are the power-law fitting curves with $c$ and $\lambda$ listed in Table I.

The power-law fitting curves are drawn in Fig. 5 and Fig. S4 in the Supplemental Material [29] with the fitting coefficient $c$ and exponent $\lambda$ being listed in Table I. It can be seen that the DFT results for all the 2D crystals can be primely fitted by the power-law with exponents around 2. As the similar power law has been found for the bending Poisson effect in 2D crystals [13], it can be expected that there should be a linear relationship between the bending induced tensile strain and bending induced axial strain in most of the 2D crystals. As the BP monolayer can become instable under strong bending deformation [16], the $\varepsilon_m \sim k$ curves deviate from the power law at large bending curvature, but still can be primely fitted by the power-law at small to moderate curvature as shown in Fig. 2(c). DFT molecular dynamic simulations at 300 K show that the bending induced extension is only slightly smaller at room temperature than that at 0 K (Supplemental Materials Fig. S5 [29]). It is also found that the bending rigidity or modulus may be over-estimated without considering the accordion effect. For example, the obtained bending moduli of monolayered BP from our DFT results are 1.61 and 6.01 eV along the armchair and zigzag directions, respectively [Fig. 2(d)], but the values become 4.07 and 8.92 eV when the bending induced extension is constrained to zero (Supplemental Materials Fig. S6 [29]).

In conclusion, we revealed a novel intrinsic mechanical behavior of 2D crystals, bending induced extension or the accordion effect. Our comprehensive DFT-based first-principles calculations show that it is the specific monolayer bonding structures and the localization of density of electron in the compressive side of the monolayer constrains the compressive deformation that result in the overall tensile deformation perpendicular to the bending moment.

This accordion effect is intrinsic to all of the investigated 2D crystals. The accordion effect can be primely described by the simple power law with an exponent around 2, making the 2D crystals more promising for novel applications, especially for mechano-electro-optical devices.

The research presented here is supported by the 973 programs (No. 2012CB937500, 2012CB933403, 2013CB932604), NSF of China (No. 51535005, 51472117, 11021262, 11172303, 11132011), Fundamental Research Funds for the Central Universities (No. NP2013309). We thank XF Liu for helpful discussions and WH Tang for help in drawing some of the geometry pictures.

[†] These authors contributed equally to this work.
*wlguo@nuaa.edu.cn

[1] K. S. Novoselov, V. I. Fal'ko, L. Colombo, P. R. Gellert, M. G. Schwab, and K. Kim, Nature **490**, 192 (2012).
[2] A.C. Ferrari, *et al.*, Nanoscale **7**, 4598 (2015).
[3] K. S. Kim, *et al.*, Nature **457**, 706 (2009).
[4] N. Kerszberg and P. Suryanarayana, RSC Adv. **5**, 43810-43814 (2015).
[5] T. Georgiou, *et al.*, Nat. Nanotech. **8**, 100 (2013).
[6] A. M. Jones, H. Yu, J. S. Ross, P. Klement, N. J. Ghimire, J. Yan, D. G. Mandrus, W. Yao, and X. Xu, Nat. Phys. **10**, 130-134 (2014).
[7] L. Li, Y. Yu, G. J. Ye, Q. Ge, X. Ou, H. Wu, D.Feng, X. H. Chen, and Y. Zhang, Nat. Nanotech. **9**, 372-377 (2014).
[8] V. M. Pereira, A. H. CastroNeto, H. Y. Liang, and L. Mahadevan, Phys. Rev. Lett. **105**, 156603 (2010).
[9] N. N. Klimov, S. Jung, S. Zhu, T. Li, C. A. Wright, S. D. Solares, D. B. Newell, N. B. Zhitenev, and J. A. Stroscio, Science **336**, 1557-1561 (2012).
[10] W. S. Bakr, A. Peng, M. E. Tai, R. Ma, J. Simon, J. I. Gillen, S. Foiling, L. Pollet, and M. Greiner, Science **329**, 544-547 (2010).
[11] C. Lee, X. Wei, J. W. Kysar, and J. Hone, Science **321**, 385-388 (2008).
[12] S. Bertolazzi, J. Brivio, and A. Kis, ACS Nano **5**, 9703-9709 (2011).
[13] R. Z. Ma, Y. Bando, and T. Sasaki, J. Phys. Chem. B **108**, 2115-2119 (2004).
[14] X. Liu, D. Pan, Y. Hong, and W. Guo, Phys. Rev. Lett. **112**, 205502 (2014).
[15] X. Zhao, Y. Liu, S. Inoue, T. Suzuki, R. O. Jones, and Y. Ando, Phys. Rev. Lett. **92**, 125502 (2004).
[16] Z. Zhang, W. Guo, and Y. Dai, Appl. Phys. Lett. **93**, 3108 (2008).
[17] D. Pan, T-C Wang, and W. Guo, Chin. Phys. B **24**, 086401 (2015).
[18] X. Shi, B. Peng, N. M. Pugno, and H. Gao, Appl. Phys. Lett. **100**, 191913 (2012).
[19] E. Cadelano, S. Giordano, and L. Colombo, Phys. Rev. B **81**, 144105 (2010).
[20] P. Koskinen, Appl. Phys. Lett. **99**, 013105 (2011).
[21] W. Kohn and L. J. Sham, Phys. Rev. **140**, A1133-A1138 (1965).
[22] G. Kresse and J. Furthmuller, Phys. Rev. B **54**, 11169 (1996).
[23] G. Kresse and J. Furthmuller, Comput. Mater. Sci. **6**, 15-50 (1996).
[24] J. P. Perdew, K. Burke, and M. Ernzerhof, Phys. Rev. Lett. **77**, 3865 (1996).
[25] P. E. Blochl, Phys. Rev. B **50**, 17953-17979 (1994).
[26] G. Kresse and J. Hafner, Phys. Rev. B **47**, 558 (1993).
[27] G. Kresse and D. Joubert, Phys. Rev. B **59**, 1758 (1999).
[28] The energy and force convergence criterias are set to be $10^{-4}$ eV and 0.01 eV/Å, respectively. The kinetic energy cutoff for the plane wave basis set is adopted to be 450 eV which guarantees the absolute energies being converged to around 2 meV. The reciprocal space for the unit cell of BP is meshed at $14 \times 10 \times 1$ using the Monkhorst Pack meshes centered at Γ point. For one-dimensional periodic cells, a vacuum space of 2 nm is included for all tubes and a dense K-point mesh ($1 \times 1 \times 12$) is used for Brillouin zone sampling.
[29] See Supplemental Material for the determination of the lengths of outer and inner half-layers of bent BP monolayers; the bending induced asymmetrical partial charge distribution; the bending to tensile deformation in other transition metal dichalcogenides; influence of finite temperature on the bending deformation and bending rigidity of monolayered BP with fixed mid-plane length.
[30] J. N. Reddy, *Theory and Analysis of Elastic Plates and Shells* (CRC Press, 2006), 2nd ed.